# Experimental realization of purely excitonic lasing in ZnO microcrystals at room temperature: transition from exciton-exciton to exciton-electron scattering

Ryosuke Matsuzaki, Haruka Soma, Kanae Fukuoka, and Takashi Uchino[*]

*Department of Chemistry, Graduate School of Science, Kobe University, 1-1 Rokkodai, Nada, Kobe 657-7501, Japan*

*email: uchino@kobe-u.ac.jp

**Abstract**

Since the seminal observation of room-temperature laser emission from ZnO thin films and nanowires, numerous attempts have been carried out for detailed understanding of the lasing mechanism in ZnO. In spite of the extensive efforts performed over the last decades, the origin of optical gain at room temperature is still a matter of considerable discussion. We show that ZnO microcrystals with a size of a few micrometers exhibit purely excitonic lasing at room temperature without showing any symptoms of electron-hole plasma emission. We then present the distinct experimental evidence that the room-temperature excitonic lasing is achieved not by exciton-exciton scattering, as has been generally believed, but by exciton-electron scattering. As the temperature is lowered below ~150 K, the lasing mechanism is shifted from the exciton-electron scattering to the exciton-exciton scattering. We also argue that the ease of carrier diffusion plays a significant role in showing room-temperature excitonic lasing.

Among other lasing materials, ZnO is one of the most well-studied optical semiconductors because of its relatively large exciton binding energy $E_b$ of about 60 meV [1,2]. Previously, there have been a number of studies on room-temperature laser emission from ZnO nanostructures, such as thin films [3–7], nanowires [8–11], nanodisks [12–14] and nanoparticles [15–18]. Two main mechanisms are generally invoked as responsible for optical gain: excitonic and electron-hole plasma (EHP) recombinations [1,2]. An EHP state is formed when the density of electron-hole pairs exceeds the Mott density $n_M$, where screening reduces the Coulomb interaction sufficiently that no bound excitonic states are present. Room-temperature EHP emission is observed commonly in most ZnO nanostructures under sufficiently high optical excitation [4–7,11,17,18]. On the other hand, room-temperature excitonic lasing is achieved almost exclusively in high-quality ZnO thin films [4–7] despite of the fact that the exciton biding energy ($E_b$ = ~60 meV) is larger than the thermal energy at room temperature (~25 meV). Note also that the exciton dominated regime generally overlaps with the EHP dominated regime [4–7]. This is most likely because rather high excitation density, which will result in the electron-hole density close to $n_M$, is required for excitonic laser action to compensate the heavy optical loss inherent to these nanostructures. Thus, realization of pure excitonic lasing at room temperature would be difficult for ZnO nanostructured materials, as has been proved in the case of ZnO nanowires [11]. To make matters more challenging, the origin of the excitonic gain in ZnO at room temperature is still controversial. Although the room-temperature excitonic lasing was originally attributed to the exciton-exciton (ex-ex) scattering process [3–7], Klingshirn and his coworkers [19,20] have recently argued that other mechanisms similar to those based on the exciton-LO phonon (ex-LO) or exciton-electron (ex-el) scattering processes are more likely to be responsible for the optical gain at room temperature. Thus, it is still an open question whether the excitonic process, especially the ex-ex scattering process, indeed accounts for the room-temperature lasing in ZnO or not [9,21].

In order to shed new light on the excitonic lasing process in ZnO, we here employ well-annealed, or well-crystallized, micrometer-sized ZnO crystals. Although lasing characteristics of the ZnO microcrystals have hardly been paid attention to previously [22–24], we show that these microcrystals have some advantages over the well-studied nanostructured materials to identify the origin of the excitonic optical gain. First, purely excitonic lasing is achieved in the ZnO microcrystals because of their high crystallinity and low optical loss. Second, individual ZnO microcrystals can serve as effective resonators to show random laser action because their size is larger than the emission wavelength [25]. Although the resulting random feedback is incoherent, the lasing line will represent the maximum of the net gain of the medium [25,26]. That is, the emission spectrum narrows continuously towards the center of the amplification line with increasing pumping intensity. As for the coherent random laser based on ZnO nanostructures [15–17], it is not straightforward to determine the position of the gain maximum on account of the presence of multiple resonance modes. From a detailed analysis of the temperature dependent lasing spectra of the ZnO microcrystals, we present the convincing experimental evidence that the excitonic lasing in ZnO at room temperature is not induced by the ex-ex scattering but by the ex-el scattering. We also demonstrated that the transition from ex-ex to ex-el process occurs at ~150 K.

We prepared micrometer-sized ZnO particles by thermal decomposition of zinc acetate dihydrate [27,28]. We first pre-annealed ~2.6 g of reagent grade zinc acetate dihydrate (Sigma-Aldrich, purity ≥98 %), which were placed in an aluminum crucible covered with an alumina lid, in an electric furnace at 300 °C for 12 h in air atmosphere. The resulting materials were washed with distilled water and dried in an oven at 90 °C for 12 h. The dried samples were then post-annealed at 800 or 1100 °C for 3 h in air, resulting in white ZnO powders. Powder X-ray diffraction (XRD) patterns were obtained with a diffractometer (Rigaku, SmartLab) using Cu Kα radiation. Scanning electron microscopy (SEM) was conducted with a scanning electron microscope (JEOL, JSM-5510). PL measurements were carried out with a gated image intensified charge-coupled device (Princeton Instruments, PI-MAX:1024RB) and 1800 or 300 lines/mm grating by using the third harmonic (355 nm) of a Q-switched Nd:yttrium aluminum garnet (YAG) laser (Spectra Physics, INDI 40, pulse width ~10 ns, repetition rate 10 Hz) as an excitation source. During the PL measurements, the laser pulse was irradiated on to a quartz sample holder containing a ZnO powder without focusing the laser beam (beam spot size of ~7 mm), and the emission signal from the front surface was monitored. The sample temperature was controlled in an optical cryostat system in the temperature region from 3 to 300 K.

Figure 1 shows the XRD patterns and SEM images of the samples prepared at temperatures of 800 and 1100 °C. Although both the samples have the hexagonal wurtzite structure of ZnO [Fig. 1(a)], the crystalline morphology varies from a deformed rod-like (~100 nm in diameter, ~2 μm in length) to an irregular spherical shape (typical size ~2 μm) depending on the preparation temperature. Room-temperature PL spectra of these micrometer-sized ZnO crystals for increasing optical excitation are shown in Fig. 2. As shown in the left panels in Fig. 2, the near-band edge (NBE) emission of the 800 °C sample shows an almost linear increase in intensity and bandwidth with increasing excitation fluence $I_{exc}$ up to ~10 mJ/cm$^2$, followed by intensity saturation at higher excitation fluences. However, the NBE emission of the 1100 °C sample exhibits a laser-like behavior upon optical pumping (see the right panels in Fig. 2); that is, the PL band at ~3.17 eV shows a nonlinear increase in the integrated emission intensity and a substantial narrowing with increasing $I_{exc}$. Considering that the ZnO crystals prepared at 1100 °C are characterized by an irregular spherical shape with a size of ~2 μm, we suggest that the observed laser-like behavior results from the random laser emission with incoherent (intensity) feedback within the respective particles, as mentioned earlier. As for the 800 °C sample, such intensity feedback will not occur in the respective crystalline rods because of the highly deformed morphology, whereas an ideally spherical ZnO microparticle exhibits discrete and coherent lasing modes due to a whispering-gallery type optical cavity [23]. Thus, the lasing behavior of the ZnO microcrystals will depend strongly on their morphology and crystallinity.

We also found that the emission peak energy for the 1100 °C sample is almost constant as $I_{exc}$ increases from ~10 to ~50 mJ/cm$^2$ [see the right panel in Fig. 2(c)]. This indicates that bandgap renormalization due to many-body effects in an EHP [29] hardly occurs under the present excitation condition. The absence of an EHP implies that the maximum carrier density attained in the 1100 °C ZnO microcrystals is below $n_M$ ~$10^{18}$ cm$^{-3}$ [11,19,30]. To confirm this implication, we estimated the density of electron-hole pairs $n_p$ as follows. Assuming that every absorbed photon creates one electron hole pair, $n_p$ can be estimated by [19]

$$n_p = \frac{P_{exc}\tau}{\hbar\omega_{ext}l}, \qquad (1)$$

where $P_{exc}$ and $\hbar\omega_{ext}$ the excitation power (in W/cm$^2$) and the photon energy (in J) of the light source, respectively. Under the quasi-stationary excitation regime, which is attained under nanosecond-pulsed excitation as employed in this work, the characteristic time $\tau$ can be considered as the decay time of the relevant emission process (spontaneous or stimulated emission). It has been demonstrated that the decay times of the spontaneous and stimulated emissions in ZnO are several hundreds of ps and a few ps, respectively [31,32]. The definition of characteristic length $l$ varies depending on whether the sample size $d$ is larger or smaller than the diffusion length $l_D$ of the excited carriers. That is,

$$l = l_D \text{ for } d > l_D, \qquad (2)$$

$$l = d \text{ for } d < l_D, \qquad (3)$$

if the effects of the carrier recombination at surface defects [19] and of the inhibition of carrier diffusion at grain boundaries [33] are neglected. As for wide direct-gap semiconductors with high crystallinity, $l_D$ can be as large as ~3 μm [29]. Thus, it can safely be assumed that in our well-annealed (1100 °C) ZnO microcrystals, $l$ is ~2 μm, which is a typical size of the microparticles. For an excitation wavelength of 355 nm with a pulse duration of 10 ns and $I_{exc} = 50$ mJ/cm$^2$, i.e., $P_{exc}=5$ MW/cm$^2$, $n_p$ is estimated to be ~5×10$^{17}$ cm$^{-3}$ on the condition that $\tau$ is 10 ps, which is an upper limit of the decay time of the stimulated emission [32]. Thus, under the excitation condition employed in this work, $n_p$ in the ZnO microcrystals will not exceed $n_M$ (~$10^{18}$ cm$^{-3}$) [34].

We next measured the excitation-fluence dependent PL spectra of the 1100 °C sample at 3 K [Fig. 3(a)]. The PL spectrum obtained under the lowest excitation fluence ($I_{exc}$=0.1 mJ/cm$^2$) shows two emission lines at 3.360 and ~3.32 eV, which can be attributed to a donor bound exciton (DBE) emission and its two electron satellite, respectively [35]. At $I_{exc}$ = ~3 mJ/cm$^2$, a new emission peak emerges on the low-energy side of the DBE emission [see the peaks indicated by arrows in Fig. 3(a)]. This peak shows a slight red shift from 3.331 to 3.313 eV with a further increase in $I_{exc}$ [Fig. 3(b)]. The integrated emission intensity of the total emission shows a nonlinear increase with increasing excitation fluence [Fig. 3(c)]. This newly emerged peak is attributed to the stimulated emission induced by ex-ex scattering, in which one exciton recombines radiatively while the other is scattered into a higher state ($n$= 2,3, ...∞). The emission maxima of the ex-ex process can be given by [2,19,36]

$$\hbar\omega_{max}^{ex-ex}(T) = E_{ex}(T) - E_b\left(1 - \frac{1}{n^2}\right) - \frac{3}{2}k_B T, \quad (4)$$

where $E_{\text{ex}}(T)$ is the free exciton transition energy at a temperature $T$, $E_{\text{b}}$ is the binding energy of the exciton, and $k_B$ is the Boltzmann constant. According to Eq. 4, the emission maxima at 3 K for $n=2$ and $\infty$ are estimated to be 3.332 and 3.316 eV, respectively, on the condition that the $E_{\text{ex}}(T)=3.377$ eV at 3 K [37]. These estimated energies for $n=2$ and $\infty$ (3.332 and 3.316 eV) are in reasonable agreement with the upper (3.331 eV) and lower (3.313 eV) limits of the observed peak maxima [see Fig. 3(b)].

To further characterize the lasing features in the 1100 °C sample at different temperatures, we measured the temperature dependent PL spectra under a constant excitation fluence of $I_{\text{exc}} = 36$ mJ/cm$^2$, which is sufficient enough to induce stimulated emission even at room temperature. As shown in Fig. 4(a). the emission band shifts to lower energies especially at temperatures above ~100 K. In Fig. 4(b), we show the temperature dependence of the observed peak maximum $\hbar\omega_{\text{max}}^{\text{obs}}(T)$ along with that of the $E_{\text{ex}}(T)$ transition energy observed for bulk ZnO [37]. One sees from Fig. 4(b) that with increasing temperature, $\hbar\omega_{\text{max}}^{\text{obs}}(T)$ shifts faster to lower energies than $E_{\text{ex}}(T)$. To more clearly show this behavior, we show in Fig. 4(c) the energy difference $\Delta E$ between $E_{\text{ex}}(T)$ and $\hbar\omega_{\text{max}}^{\text{obs}}(T)$. For comparison, we also plot the energy difference between $E_{\text{ex}}(T)$ and $\hbar\omega_{\text{max}}^{\text{ex-ex}}(T)$, i.e., $E_{\text{b}}^{\text{ex}}\left(1-\frac{1}{n^2}\right)+\frac{3}{2}k_B T$ [see Eq. (4)], for $n=2$ and $\infty$ [see the broken lines in Fig. 4(c)]. One sees that at temperatures from 3 to ~150 K, $\Delta E$ is located in the energy range between the lines calculated for $n=2$ and $\infty$, indicating that the ex-ex process persists in this temperature regime. At temperatures above ~150 K, however, $\Delta E$ is not located within the energy range of the ex-ex process but shows a clear linear dependence with temperature with a slope of $4.78\times10^{-4}$ eV/K. We also confirmed that the value of slope is hardly changed even when the excitation fluence is increased to ~ ~50 mJ/cm$^2$. These results clearly indicate that the ex-el process prevails in the temperature region from ~150 to 300 K because the emission maximum of the ex-el process is given by [2,29,36]

$$\hbar\omega_{\text{max}}^{\text{ex-el}}(T) = E_{\text{ex}}(T) - \gamma k_\text{B} T, \qquad (5)$$

where $\gamma$ is a constant related to the ratio of exciton effective mass over electron effective mass [36]. The slope of $4.78\times10^{-4}$ eV/K yields the value of $\gamma= 5.6$, which is in reasonable agreement with the predicted value of $\gamma =$~7 [36]. Thus, we conclude that the room-temperature lasing peak at ~3.17 eV observed in the 1100 °C sample is attributed not to the ex-ex process, as has often been believed [3-7], but to the ex-el process. In Supplemental Material [38], we also provide a reason for the misinterpretation of the room-temperature lasing peak at ~3.17 eV observed previously in ZnO thin films.

The above investigations and considerations support the determinant role of the crystalline quality on the lasing properties of ZnO – in terms not only of optical loss but also of carrier diffusion. A higher (lower) crystalline quality will lead to a longer (shorter) diffusion length of carriers, resulting in a decrease (increase) in the practical carrier density during photoexcitation [33]. This can explain why the excitonic lasing has been observed exclusively in high-quality ZnO thin films. For low-quality ZnO thin films, the electron-hole density could soon reach the level of $n_M$ during photoexcitation because of inhibition of exciton diffusion, hence resulting in an EHP state. Thus, for efficient realization of excitonic lasing, one should use the high crystallinity microcrystals, as in the case of the present 1100 °C sample, in which the total exciton density is to be distributed over the sample to avoid reaching $n_M$ within the timescale of exciton recombination.

In summary, we have shown that the sphere-like shaped ZnO microcrystals are useful for the investigation of the gain mechanism in ZnO. We have reported the direct experimental demonstration of the existence of the ex-el process at room temperature. More generally, the present observations have presented the distinct experimental evidence of the transition from the ex-el lasing to the ex-ex lasing at ~150 K. We have also argued that the size and crystallinity of ZnO play a vital role in showing pure excitonic lasing at room temperature in terms of exciton diffusion.

[34] However, this argument will not valid for the 800 °C sample because of the lower

crystallinity, which will lead to a shorter value of $l$ and hence a higher value of $n_\text{p}$.

The red shift and spectral broadening observed in the 800 °C sample is likely to be indicative of an EHP state although lasing is not achieved because of the expected heavy optical loss and the absence of effective feedback.

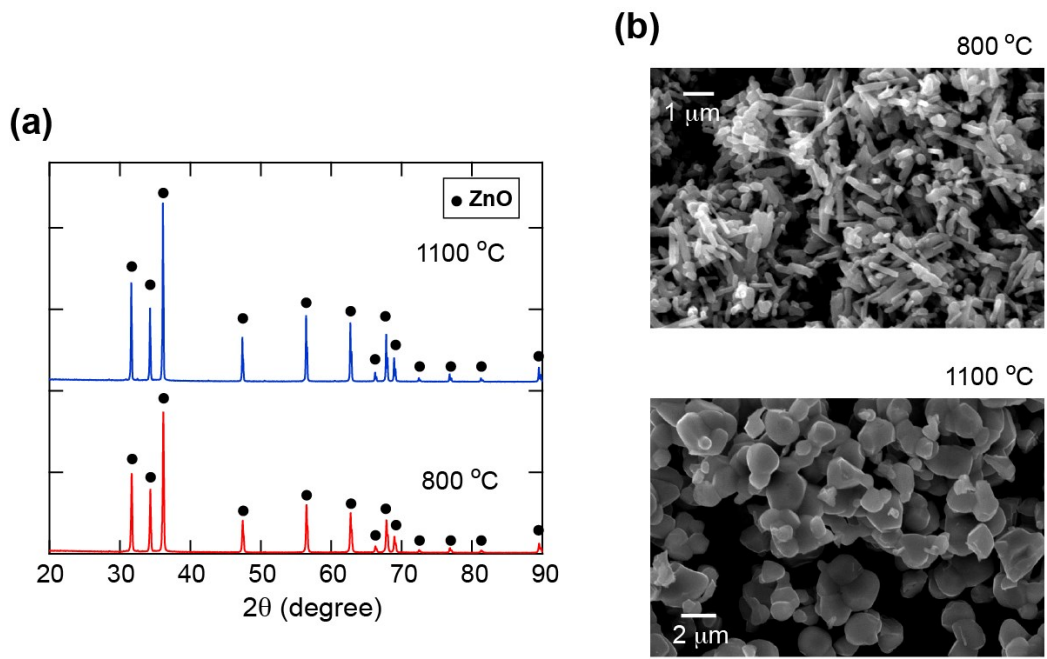

FIG. 1. (a) XRD patterns and (b) SEM images of the ZnO microcrystals. The samples were prepared by thermal decomposition of zinc acetate dihydrate at temperatures of 800 and 1100 °C in air atmosphere.

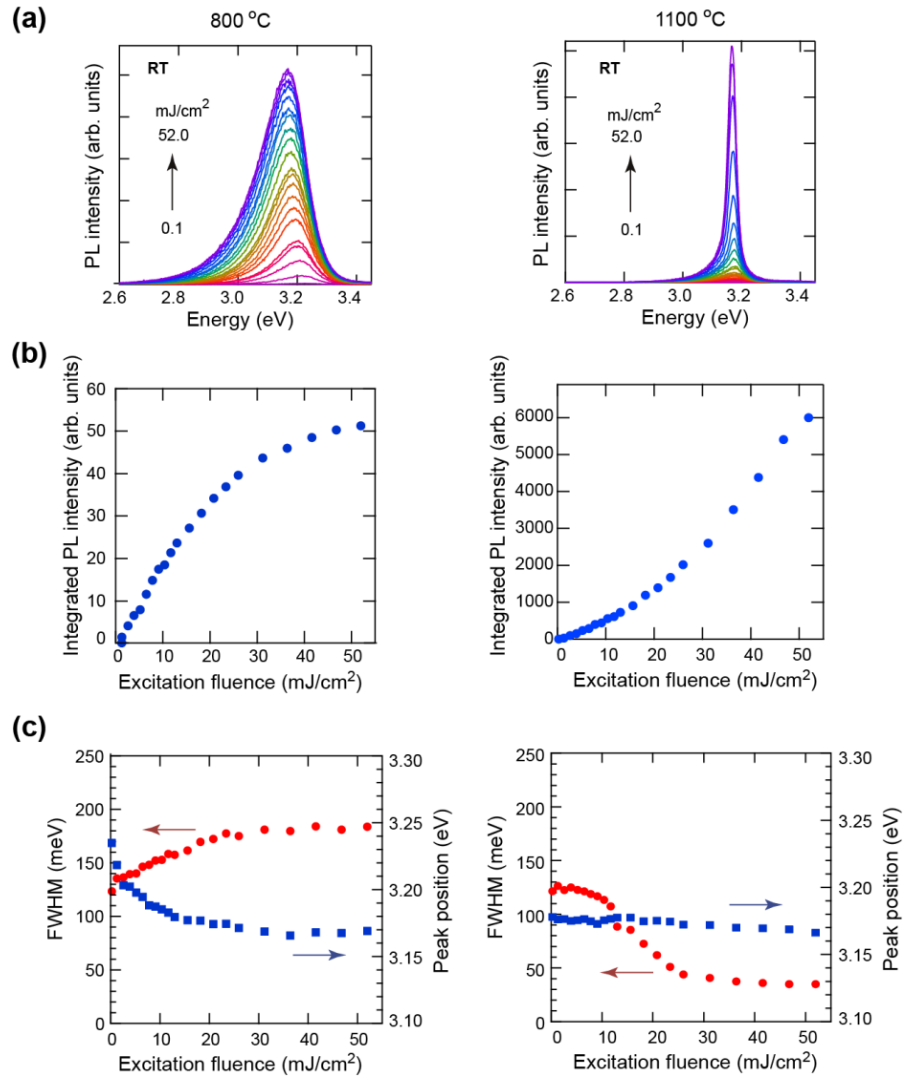

FIG. 2. Excitation fluence dependence of the room-temperature PL characteristics of the ZnO microcrystals. Left and right panels show the results for the ZnO microcrystals prepared at 800 and 1100 °C, respectively. (a) Changes in the PL spectra with increasing excitation fluence from 0.1 to 52 mJ/cm$^2$. The third harmonic ($\lambda$=355 nm) of a 10-ns pulsed Nd:YAG laser was used as an excitation source. (b) Excitation fluence dependence of the spectrally-integrated PL intensity. (c) Excitation fluence dependence of the full width at a half maximum (FWHM, left axis) and the PL peak position (right axis).

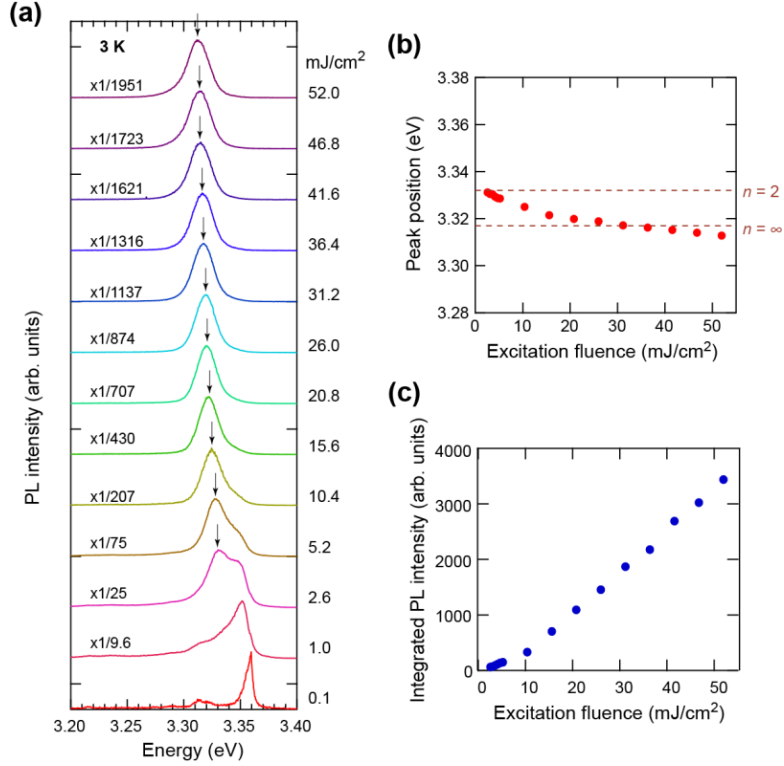

FIG. 3. PL characteristics of the 1100 °C prepared ZnO microcrystals measured at 3 K.
(a) Changes in the PL spectra with increasing excitation fluence from 0.1 to 52 mJ/cm$^2$. The peak intensities are normalized among all spectra, which are displaced vertically for clarity. Excitation fluence dependence of (b) the peak energy marked by arrows in Fig. 2a and (c) the spectrally-integrated PL intensity. In (b), the emission maxima of the ex-ex process for $n=2$ and $\infty$ at 3 K calculated from Eq. (4) 1 are indicated by red dashed lines.

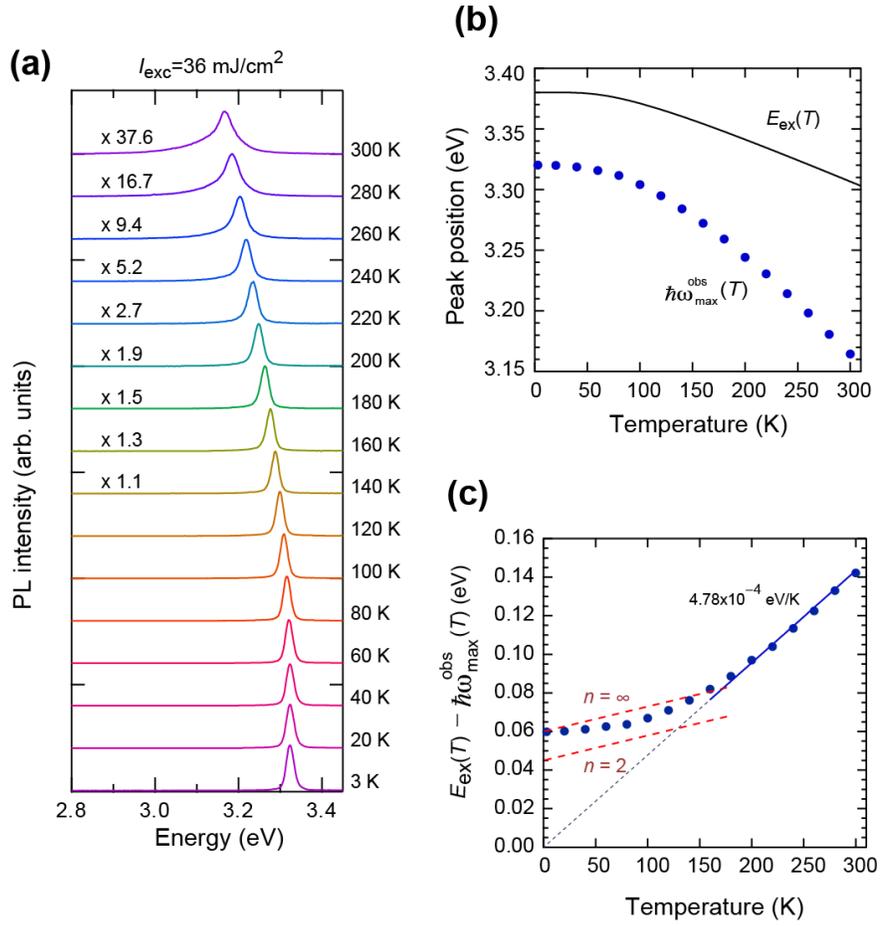

FIG. 4 Temperature dependence of the lasing peak observed for the 1100 °C prepared ZnO microcrystals. (a) Changes in the lasing spectra with increasing temperature from 3 to 300 K measured under a constant excitation fluence of 36 mJ/cm². The peak intensities are normalized among all spectra, which are displaced vertically for clarity. (b) The peak energy $\hbar\omega_{max}^{obs}$ as a function of temperature. The free-exciton transition energy $E_{ex}$ reported in [37] is also shown as a solid line. (c) The energy difference between $E_{ex}$ and $\hbar\omega_{max}^{obs}$ as a function of temperature. The solid line shows a least squares fit of the data in the temperature region from 160 to 300 K to Eq. (5). The energy differences between $E_{ex}(T)$ and $\hbar\omega_{max}^{ex-ex}(T)$ for $n=2$ and $\infty$ calculated from Eq. (4) are also shown as red dashed lines.